\documentclass[pra,aps,amsmath,amssymb,amsfonts,twocolumn,nofootinbib,showpacs,longbibliography]{revtex4-1}

\usepackage{amssymb}
\usepackage{}
\usepackage{bm,mathrsfs}

\usepackage{graphicx}
\usepackage{epsfig}
\usepackage{amsmath,bbm}
\usepackage{amsfonts,amssymb}
\usepackage{times}
\usepackage{verbatim}
\usepackage[sort&compress]{natbib}

\usepackage{amsmath}
\usepackage[colorlinks,breaklinks,linkcolor=blue,anchorcolor=blue,citecolor=blue,urlcolor=blue,
dvipdfm
]{hyperref}

\newcommand{\be}{\begin{equation}}
\newcommand{\ee}{\end{equation}}
\newcommand{\bea}{\begin{eqnarray}}
\newcommand{\eea}{\end{eqnarray}}

\def\>{\rangle}
\def\<{\langle}

\def\qed{\leavevmode\unskip\penalty9999 \hbox{}\nobreak\hfill
     \quad\hbox{\leavevmode  \hbox to.77778em{%
               \hfil\vrule   \vbox to.675em%
               {\hrule width.6em\vfil\hrule}\vrule\hfil}}
     \par\vskip3pt}
     
\begin{document}

\newtheorem{theorem}{Theorem}
\newtheorem{lemma}[theorem]{Lemma}
\newtheorem{corollary}[theorem]{Corollary}
\newtheorem{proposition}[theorem]{Proposition}
\newtheorem{definition}[theorem]{Definition}
\newtheorem{example}[theorem]{Example}
\newtheorem{conjecture}[theorem]{Conjecture}
\title{Unconventional Rydberg pumping and applications in quantum information processing}

\author{D. X. Li}
\affiliation{Center for Quantum Sciences and School of Physics, Northeast Normal University, Changchun, 130024, People's Republic of China}
\affiliation{Center for Advanced Optoelectronic Functional Materials Research, and Key Laboratory for UV Light-Emitting Materials and Technology
of Ministry of Education, Northeast Normal University, Changchun 130024, China}

\author{X. Q. Shao}
\email{Corresponding author: shaoxq644@nenu.edu.cn}
\affiliation{Center for Quantum Sciences and School of Physics, Northeast Normal University, Changchun, 130024, People's Republic of China}
\affiliation{Center for Advanced Optoelectronic Functional Materials Research, and Key Laboratory for UV Light-Emitting Materials and Technology
of Ministry of Education, Northeast Normal University, Changchun 130024, China}


\date{\today}

\begin{abstract}
{We propose a mechanism of unconventional Rydberg pumping (URP) via simultaneously driving each Rydberg atom by two classical fields with different strengths of Rabi frequencies. This mechanism differs from the general Rydberg blockade or Rydberg antiblockade since it is closely related to the ground states of atoms, i.e. two atoms in the same ground state are stable while two atoms in different ground states are resonantly excited. Furthermore, we find the URP can be employed to simplify some special quantum information processing tasks, such as implementation of a three-qubit controlled phase gate with only a single Rabi oscillation, preparation of two- and three-dimensional steady-state entanglement with two identical atoms, and realization of the autonomous quantum error correction in a Rydberg-atom-cavity system. The feasibility of the above applications is certified explicitly by the state-of-the-art technology.}
\end{abstract}

\pacs {03.67.Bg, 03.65.Yz, 32.80.Qk, 32.80.Ee}\maketitle \maketitle

\section{Introduction}\label{I}
The unique feature of the interatomic Rydberg interactions opens many possibilities to explore neutral atoms in the researches of few- and many-body physics and quantum information applications \cite{pra042306ref1}. One of the critical effects is the Rydberg blockade: In a small volume, once a Rydberg atom is excited to the Rydberg state, the strong, long-range interactions between Rydberg atoms will significantly suppress the other Rydberg atoms excited. After the first scheme to perform fast gate operations by the Rydberg blockade was proposed by Jaksch \textit{et al.} \cite{PhysRevLett.85.2208}, a variety of proposals were designed theoretically and experimentally for quantum computation \cite{pra052313ref34,pra052313ref36,pra052313ref38,PhysRevApplied.7.064017,PhysRevApplied.9.051001}, entanglement generation \cite{pra052313ref39,pra012328ref8,pra052313ref41,pra052313ref42,PhysRevA.97.032701},  quantum algorithms \cite{pra052313ref43}, quantum simulators \cite{pra052313ref44}, and quantum repeaters \cite{pra052313ref45}. {Another dramatic effect making use of the interatomic Rydberg interactions is the Rydberg dressing, which results from the adiabatical dressing between the ground state and the
excited Rydberg state \cite{nphys3487ref25,nphys3487ref26,nphys3487ref27}. It enables tunable, anisotropic interactions and provides the possibility to study the novel exotic many-body physics   \cite{pra043606ref20,nphys3487ref29,pra043606ref22,pra043606ref24,nphys3487}}

In addition, the combination of interatomic Rydberg interactions and two-photon detuning leads to an opposite effect, the Rydberg antiblockade, which theoretically predicted by Ates \textit{et al.} \cite{PhysRevLett.98.023002} and was experimentally observed by Amthor \textit{et al.} \cite{PhysRevLett.104.013001}. The Rydberg antiblockade can achieve the multiple
Rydberg excitations while restrain the Rydberg blockade, and its exploration is of particular interest not only for multiqubit
logic gates \cite{pra022319}, but also for preparations of quantum entanglement  \cite{pra012328ref20,PhysRevA.92.022328,pra032336,PhysRevA.96.062315,Li:18}, \textit{e.g.} {Carr \textit{et al.} analyzed an approach to obtain high fidelity entanglement and antiferromagnetic states by Rydberg antiblockade \cite{pra012328ref20}.} Quite recently, our group made use of the cooperation between Rydberg antiblockade, quantum Zeno dynamics, and atomic spontaneous emission to prepare the tripartite GHZ state and $W$ state, respectively \cite{PhysRevA.96.062315,Li:18}. {Moreover, by virtue of the Rydberg-antiblockade effect and the Raman transition, we have devised a mechanism of ground-state blockade to generate the high-fidelity entanglement \cite{pra012328}.}

In this paper, we propose an unconventional Rydberg pumping (URP), which is different from the all above effects. This effect is closely related to the ground states of atoms, i.e. two atoms in the same ground state are stable while two atoms in different ground states are resonantly excited. Taking the case of two atoms as an example, the corresponding atomic level has been shown in Fig.~\ref{gate}(a). Both of atoms consist of two ground states $|0\rangle$ and $|1\rangle$, and one Rydberg state $|r\rangle$. For each atom, the Rydberg state is dispersively coupled with the ground state $|1\rangle$ via a classical field of Rabi frequency $\Omega_1$, detuning $-\Delta$, and another classical field resonantly drives the transition $|1\rangle\leftrightarrow|r\rangle$ with Rabi frequency $\Omega_2$. Exploiting the URP, we can freeze the evolution of state $|11\rangle$, and only the states $|10\rangle$ and $|01\rangle$ can be excited into $|r0\rangle$ and $|0r\rangle$, respectively. In contrast, it is impossible for the Rydberg blockade and Rydberg antiblockade to inhibit the transitions $|11\rangle\leftrightarrow(|1r\rangle+|r1\rangle)/\sqrt{2}$ or $|11\rangle\leftrightarrow|rr\rangle$ and realize $|10(01)\rangle\leftrightarrow|r0(0r)\rangle$, simultaneously. We will illustrate the mechanism detailedly in the next section and discuss its applications in three-qubit controlled phase gate, steady-state entanglement and autonomous quantum error correction in succession.

The remainder of the paper is organized as follows. In Sec.~\ref{II}, we illustrate the mechanism of the URP in detail with two three-level atoms. In Sec.~\ref{III}, we apply the new technique to achieve a three-qubit controlled phase gate. In Sec.~\ref{IV}, we discuss the possibility to dissipatively prepare the two- and three-dimensional entangled states by the URP. In Sec.~\ref{V}, we make use of the URP to realize the autonomous quantum error correction in a Rydberg-atom-cavity system. We summarize our works in Sec.~\ref{VI}.

\section{Mechanism of the URP between two atoms}\label{II}
\begin{figure}
\centering
\includegraphics[scale=0.20]{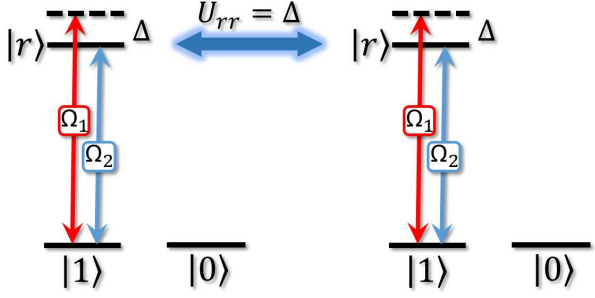}
\caption{Two atomic level configuration for the URP. The Rydberg state is dispersively coupled with the ground state $|1\rangle$ via a classical field of Rabi frequency $\Omega_1$, detuning $-\Delta$, and another classical field resonantly drives the transition $|1\rangle\leftrightarrow|r\rangle$ with Rabi frequency $\Omega_2$.}
\label{gate}
\end{figure}

{The effectiveness of the URP is not limit to the two-atom case. But it is instrumental enough for us  to interpret the mechanism of the URP clearly with a bipartite system.} The system includes two $\Lambda$-type three-level Rydberg atoms which is shown in Fig.~\ref{gate}(a). The quantum information is encoded into the subspace $\{|00\rangle, |01\rangle, |10\rangle, |11\rangle\}.$ In the interaction picture, the Hamiltonian of the system can be written as
\begin{equation}\label{H}
H_I=\sum_{i=1}^2\Omega_1 e^{-i\Delta t}|r\rangle_i\langle1|+\Omega_2|r\rangle_i\langle1|+{\rm H.c.}+U_{rr}|rr\rangle\langle rr|,
\end{equation}
where the subscript $i$ means the $i$-th atom and $U_{rr}$ denotes the Rydberg-mediated interactions. When we choose $U_{rr}=\Delta$, the Hamiltonian can be reformulated with respect of a rotated frame as
\begin{eqnarray}\label{rH}
H_I=H_I^{(1)}+H_I^{(2)}+H_I^{(3)},
\end{eqnarray}
where
\begin{eqnarray}
H_I^{(1)}&=&\Omega_2(|r0\rangle\langle10|+|0r\rangle\langle01|)+{\rm H.c.},\nonumber\\
H_I^{(2)}&=&\sqrt{2}(\Omega_1|D\rangle\langle rr|+\Omega_2|D\rangle\langle 11|)+{\rm H.c.},\nonumber\\
H_I^{(3)}&=&\sqrt{2}e^{-i\Delta t}(\Omega_1|D\rangle\langle11|+\Omega_2|D\rangle\langle rr|)\nonumber\\
&&+\Omega_1e^{-i\Delta t}(|r0\rangle\langle10|+|0r\rangle\langle01|)+{\rm H.c.},\nonumber
\end{eqnarray}
and $|D\rangle=(|1r\rangle+|r1\rangle)/\sqrt{2}$. In the limit of $\Delta\gg\Omega_1,\Omega_2$, $H_I^{(3)}$ consisting of the high-frequency oscillating terms can be neglected. Then we expand $H_I^{(2)}$ in terms of the basis of $\{|11\rangle,|+\rangle,|-\rangle  \}$, where $|\pm\rangle=(|rr\rangle\pm|D\rangle)/\sqrt{2}$ are the eigenvectors of $\sqrt{2}\Omega_1|D\rangle\langle rr|+{\rm H.c.}$ with respect to the eigenvalues $\pm\sqrt{2}\Omega_1$, i.e.
\begin{eqnarray}
H_I^{(2)}&=&\sqrt{2}\Omega_1(|+\rangle\langle+|-|-\rangle\langle-|)+\Omega_2(|+\rangle\langle11|\nonumber\\
&&-|-\rangle\langle11|+{\rm H.c.}).
\end{eqnarray}
From the above equation, we can find that as $\Omega_1\gg\Omega_2$, a quantum state initialized in $|11\rangle$ will not evolve into others since the corresponding detunings are $\pm\sqrt{2}\Omega_1$. Therefore the Hamiltonian $H_I^{(2)}$ of Eq.~(\ref{rH}) can be also neglected further.
Finally, the total Hamiltonian has been simplified as
\begin{eqnarray}\label{effH}
H_{I}\simeq H_{\rm eff}=H_I^{(1)}=\Omega_2(|r0\rangle\langle10|+|0r\rangle\langle01|)+{\rm H.c.}.
\end{eqnarray}
 During the whole derivation, the Stark-shift terms are not considered since they can be canceled via introducing other ancillary levels. Eq.~(\ref{effH}) signifies that in the limiting condition of URP, i.e. $U_{rr}=\Delta\gg\Omega_1\gg\Omega_2$, the qubits system will not evolve except for the subspace spanned by $\{ |01\rangle,|10\rangle \}$.

\begin{figure}
\centering
\includegraphics[scale=0.4]{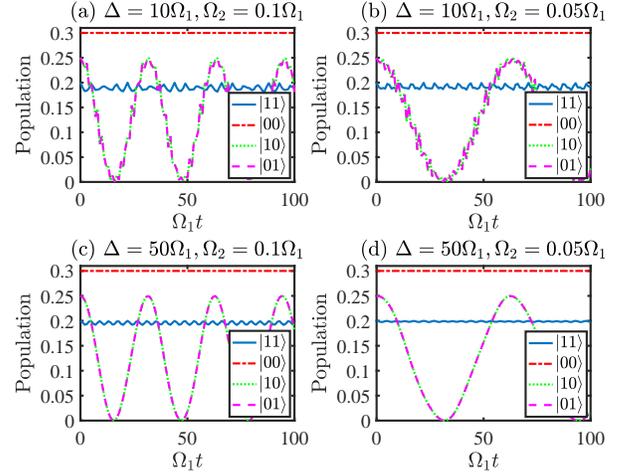}
\caption{The populations as functions of $\Omega_1 t$ governed by the Liouville equation $\dot\rho(t)=-i[H_I,\rho(t)]$ with different $\Delta/\Omega_1$ and $\Omega_1/\Omega_2$, where the population of state $|\psi\rangle$ is defined as $P=\langle\psi|\rho(t)|\psi\rangle$ and the initial states are all chosen as an random mixed state $\rho_0=0.2|11\rangle\langle11|+0.3|00\rangle\langle00|+0.25|10\rangle\langle10|+0.25|01\rangle\langle01|$.  }
\label{UD}
\end{figure}

For explicitly determining the suitable values to satisfy the limiting condition of URP, we plot the evolutions of populations of states $|11\rangle$ (solid line), $|00\rangle$ (doted-dashed line), $|10\rangle$ (dotted line) and $|01\rangle$ (dashed line) with different values of $\Delta/\Omega_1$ and $\Omega_1/\Omega_2$ in Fig.~\ref{UD}. We can find that the larger the values of $\Delta/\Omega_1$ and $\Omega_1/\Omega_2$, the more stable the state $|11\rangle$. Furthermore, $\Delta=50\Omega_1$ and $\Omega_2=0.05\Omega_1$ are good enough to excellently execute the URP as shown in Fig.~\ref{UD}(d).

\section{three-qubit controlled phase gate}\label{III}

\begin{figure}
\centering
\includegraphics[scale=0.20]{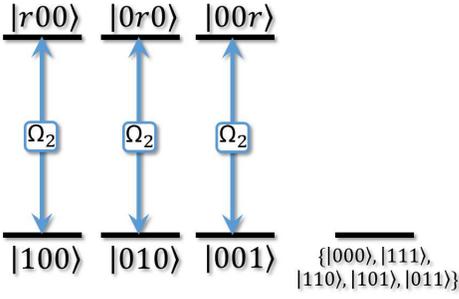}
\caption{The effective transitions of three-qubit controlled phase gate. Only the states with one atom in $|1\rangle$ will undergo a Rabi oscillation with Rabi frequency $\Omega_2$ while the other states are stable.}
\label{gateeff}
\end{figure}

\begin{figure}
\centering
\includegraphics[scale=0.52]{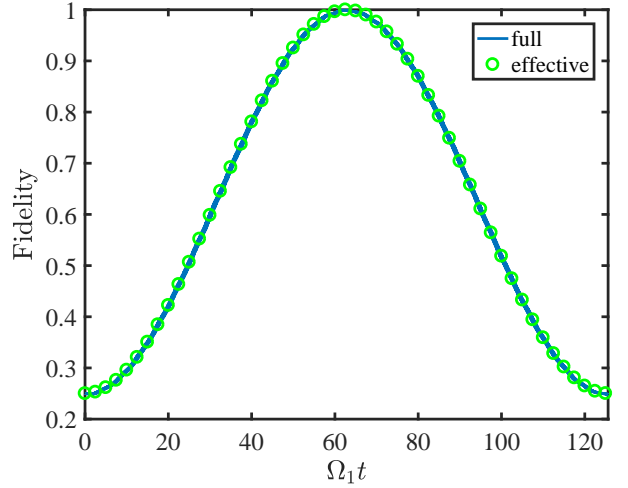}
\caption{The evolutions of corresponding fidelity governed by the $H_I$ of Eq.~(\ref{gateH}) (solid line) and the $H_{\rm eff}$ of Eq.~(\ref{gateeffH}) (empty circles), respectively. The initial state is $|\psi_0\rangle$ and the parameters are chosen as $\Omega_2=0.05\Omega_1,$ and $\Delta=58\Omega_1$.}
\label{gateF}
\end{figure}

It is universally acknowledged that the $n$-qubit controlled phase gate is an essential ingredient for quantum algorithms \cite{pra012326ref1,pra012326ref4,PhysRevLett.79.4709} and quantum Fourier transform \cite{pra012326ref5}. Here we implement a three-qubit controlled phase gate by the URP, which is fast completed by a single Rabi oscillation for a single atom. The atomic level of each atom remains the same as Fig.~\ref{gate}(a) and the Hamiltonian reads
\begin{eqnarray}\label{gateH}
H_I&=&\sum_{i=1}^3\Omega_1 e^{-i\Delta t}|r\rangle_i\langle1|+\Omega_2|r\rangle_i\langle1|+{\rm H.c.}\nonumber\\
&&+\sum_{j>i}U_{rr}|rr\rangle_{ij}\langle rr|,
\end{eqnarray}
where we have considered the interactions between different atoms are identical to $U_{rr}$.
In the limiting condition $U_{rr}=\Delta\gg\Omega_1\gg\Omega_2$, on the basis of the analysis in Sec.~\ref{II}, it is evident that the system initialized in $\{ |000\rangle,|110\rangle,|101\rangle, |011\rangle, \}$ will be stable.
What makes the tripartite system different from the bipartite system is that
for the system initialized in $|111\rangle$, the Rydberg antiblockade will result in an effective Hamiltonian $6\Omega_1^3(|111\rangle\langle rrr|+|rrr\rangle\langle111|)/\Delta^2$. Although these terms are adverse for our purpose, it can be ignored because the contribution of $6\Omega_1^3/\Delta^2$ is much smaller than $\Omega_2$ in the limiting condition $\Delta\gg\Omega_1$. Thus the total effective Hamiltonian of the three-qubit controlled gate can be written as
\begin{eqnarray}\label{gateeffH}
H_{\rm eff}=\Omega_2(|r00\rangle\langle100|+|0r0\rangle\langle010|+|00r\rangle\langle001|)+{\rm H.c.}.
\end{eqnarray}
In Fig.~\ref{gateeff}, we show the effective transitions of the three-qubit controlled phase gate according to the Eq.~(\ref{gateeffH}). In our system, only the states with one atom in $|1\rangle$ will undergo a Rabi oscillation with Rabi frequency $\Omega_2$, while the other states are stable. Consequently, we carry out the three-qubit controlled phase gate after the interaction time $T=\pi/\Omega_2$, which map the direct product state of three atoms $|\psi_0\rangle=(|0\rangle+|1\rangle)_1(|0\rangle+|1\rangle)_2(|0\rangle+|1\rangle)_3/2\sqrt{2}$ into the three-atom entanglement $|\psi_s\rangle=(|000\rangle+|011\rangle+|101\rangle+|110\rangle+|111\rangle-|100\rangle-|010\rangle-|001\rangle)/2\sqrt{2}$. To demonstrate the feasibility of our scheme, we compare the evolutions of corresponding fidelity governed by the $H_I$ of Eq.~(\ref{gateH}) (solid line) and $H_{\rm eff}$ of Eq.~(\ref{gateeffH}) (empty circles) in Fig.~\ref{gateF}, respectively. The fidelity is defined as $F=|\langle\psi_s|\exp(-iH_It)|\psi_0\rangle|$. In Fig.~\ref{gateF}, the two curves are in good agreement with each other and the corresponding fidelity can reach $99.94\%$, which adequately illuminate the validity of the effective system and the feasibility of the mechanism.
In experiment, the Rydberg atoms with suitable principal quantum number can achieve the long radiative lifetimes \cite{PhysRevA.85.065401}, which can inhibit the detriment of atom spontaneous emission for the three-qubit controlled phase gate, \textit{e.g.} the $97d_{5/2}$ Rydberg state with the decay rate $\gamma\simeq2\pi\times1$ KHz \cite{prl090402}. When we consider the Rydberg state $|r\rangle$ decays to the ground states with the same rate $\gamma/2=2\pi\times1/2$ KHz, the decay of $i$-th atom can be described as Lindblad operators $L_i^{0(1)}=\sqrt{\gamma/2}|0(1)\rangle_i\langle r|$ and the evolution of system will be governed by the master equation
\begin{eqnarray}\label{gatemaster}
\dot\rho=-i[H_I,\rho]+\mathcal{L}\rho,
\end{eqnarray}
where
\begin{eqnarray}
\mathcal{L}\rho=\sum_{i=1}^3\sum_{k=0}^1L_i^k\rho L_i^{k\dag}-\frac{1}{2}(L_i^{k\dag}L_i^k\rho+\rho L_i^{k\dag}L_i^k).
\end{eqnarray}
Meanwhile, the Rabi laser frequency $\Omega_1$ and $\Omega_2$ can be tuned continuously between $2\pi\times(0,100)$ MHz in experiment \cite{prl090402,PhysRevA.96.062315}. Thus, the other parameters are set as $(\Omega_1,\Omega_2,\Delta)=2\pi\times(1,0.05,58)$ MHz and a high fidelity $F=\sqrt{\langle \psi_s|\rho(\pi/\Omega_2)|\psi_s\rangle}=99.37\%$ can be guaranteed.

\section{dissipative generation of entanglement }\label{IV}
\subsection{Two-dimensional entangled state}

\begin{figure}
\centering
\includegraphics[scale=0.15]{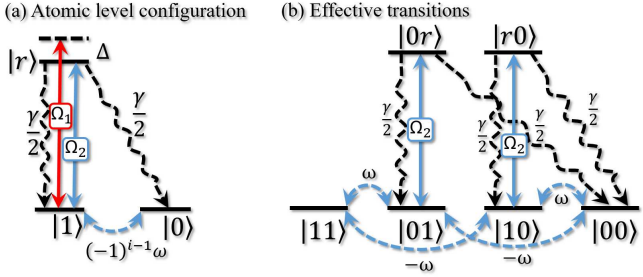}
\caption{(a) The atomic level configuration of the scheme to dissipatively prepare bipartite entanglement. (b) The corresponding effective transitions.}
\label{2Dmodel}
\end{figure}

{With the rapid development of quantum information, more and more interest has been devoted to preparing quantum entanglement with the quantum noise, which can be regarded as a resource \cite{PhysRevLett.106.090502,pra012319,PhysRevLett.117.040501}.
Combining the URP with the spontaneous emission of two Rydberg atoms, we propose a dissipative way to generate the Bell state $|\phi_+\rangle=(|00\rangle+|11\rangle)/\sqrt{2}$, which is independent of the initial state.}  As shown in Fig.~\ref{2Dmodel}(a), in addition to the classical fields driving the transition between $|r\rangle$ and $|1\rangle$ dispersively and resonantly, we add the microwave fields to resonantly drive $|1\rangle\leftrightarrow|0\rangle$ with Rabi frequency $(-1)^{i-1}\omega$, where $i=1,2$ denotes the $i$-th atom. The branching ratios of spontaneous emission for $i$-th atom from $|r\rangle$ downwards to $|0\rangle$ and $|1\rangle$ are both assumed to be $\gamma/2$, described by the Lindblad operators $L_i^{0(1)}=\sqrt{\gamma/2}|0(1)\rangle_i\langle r|$. The corresponding Hamiltonian and full master equation can be respectively indicated as
\begin{eqnarray}\label{2DH}
H_I&=&H_L+H_{MW}\\
H_L&=&\sum_{i=1}^2\Omega_1 e^{-i\Delta t}|r\rangle_i\langle1|+\Omega_2|r\rangle_i\langle1|+{\rm H.c.}\nonumber\\
&&+U_{rr}|rr\rangle\langle rr|,\nonumber\\
H_{MW}&=&\sum_{i=1}^2(-1)^{i-1}\omega|1\rangle\langle0|+{\rm H.c.},\nonumber
\end{eqnarray}
and
\begin{eqnarray}\label{2Dmaster}
\dot\rho&=&-i[H_I,\rho]+\mathcal{L}\rho,\\
\mathcal{L}\rho&=&\sum_{i=1}^2\sum_{k=0}^1L_i^k\rho L_i^{k\dag}-\frac{1}{2}(L_i^{k\dag}L_i^k\rho+\rho L_i^{k\dag}L_i^k).\nonumber
\end{eqnarray}
According to the principle of the URP, $H_L$ can be simplified as $H_L=\Omega_2(|01\rangle\langle0r|+|10\rangle\langle r0|)+{\rm H.c.}$. Then expanding the $H_{MW}$ with the basis of $\{ |11\rangle,|01\rangle,|10\rangle,|00\rangle   \}$, we can reformulate the Hamiltonian as follows
\begin{eqnarray}\label{2DeffH}
H_{\rm eff}&=&\Omega_2(|10\rangle\langle r0|+|01\rangle\langle0r|)+\omega(|11\rangle-|00\rangle)\nonumber\\
&&\otimes(\langle01|-\langle10|)+{\rm H.c.},
\end{eqnarray}
with the corresponding effective master equation
\begin{eqnarray}\label{2DeffM}
\dot\rho&=&-i[H_{\rm eff},\rho]+\mathcal{L_{\rm eff}}\rho,\\
\mathcal{L_{\rm eff}}\rho&=&\sum_{k=1}^4L_{\rm eff}^k\rho L_{\rm eff}^{k\dag}-\frac{1}{2}(L_{\rm eff}^{k\dag}L_{\rm eff}^k\rho+\rho L_{\rm eff}^{k\dag}L_{\rm eff}^k),\nonumber
\end{eqnarray}
where
\begin{eqnarray}
&&L_{\rm eff}^1=\sqrt{\frac{\gamma}{2}}|01\rangle\langle0r|,\ \ L_{\rm eff}^2=\sqrt{\frac{\gamma}{2}}|00\rangle\langle0r|,\nonumber\\
&&L_{\rm eff}^3=\sqrt{\frac{\gamma}{2}}|10\rangle\langle r0|,\ \ L_{\rm eff}^4=\sqrt{\frac{\gamma}{2}}|00\rangle\langle r0|.\nonumber
\end{eqnarray}

In Fig.~\ref{2Dmodel}(b), we illustrate the effective transitions to intuitively explain the operational principle. The interconversion between four ground states $|11\rangle,|01\rangle,|10\rangle$ and $|00\rangle$ is realized by the microwave fields, and the ground states $|10\rangle$ and $|01\rangle$ are also coupled with the excited states $|r0\rangle$ and $|0r\rangle$ which will then spontaneously decay to the ground states $|01\rangle,|10\rangle$ and $|00\rangle$. The total transitions construct a cyclic evolution of system and we can find that the Bell state $|\phi_+\rangle$ is the unique steady-state solution of Eq.~(\ref{2DeffM}) because of $H_{\rm eff}|\phi_+\rangle=L_{\rm eff}^k|\phi_+\rangle=0$. Therefore, the system will be stabilized at the state $|\phi_+\rangle$ ultimately.

\begin{figure}
\centering
\includegraphics[scale=0.52]{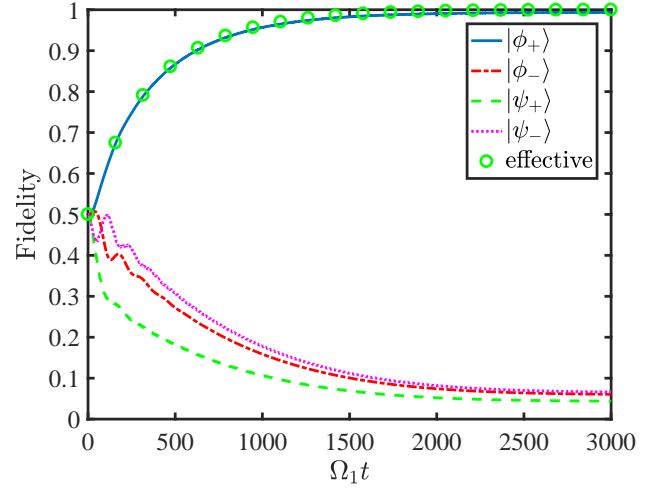}
\caption{The evolutions of fidelity for the Bell states governed by the full and effective master equation. The initial state is $\rho_0=0.25|00\rangle\langle00|+0.25|10\rangle\langle10|+0.25|01\rangle\langle01|+0.25|11\rangle\langle11|$. The relevant parameters are chosen as $\Omega_2=0.02\Omega_1,\omega=0.01\Omega_1,\Delta=100\Omega_1$ and $\gamma=0.05\Omega_1$.}
\label{2DF}
\end{figure}

In Fig.~\ref{2DF}, we compare the time evolutions of the fidelity for the target state $|\phi_+\rangle$ governed by the full master equation (solid line) and the effective master equation (empty circles) to confirm the validity of the above derivations. The tendencies of the two curves are identical, which implies that the reduced system is accurate and we can forecast the behavior of the realistic system by the reduced system. On the other hand, we find that the fidelity of the target state can arrive at $99.35\%$ and the fidelities of states $|\phi_-\rangle=(|00\rangle-|11\rangle)/\sqrt{2}$ (dashed-dotted line), $|\psi_+\rangle=(|01\rangle+|10\rangle)/\sqrt{2}$ (dashed line) and $|\psi_-\rangle=(|01\rangle-|10\rangle)/\sqrt{2}$ (dotted line) all tend to be vanished, which fully explicate the feasibility of the dissipative scheme. When we choose the experimental parameters as $(\Omega_1,\Omega_2,\omega,\Delta,\gamma)=2\pi\times(1,0.02,0.01,100,0.03)$ MHz \cite{njp043020}, the fidelity of the target state can be above $99.48\%$.
\subsection{Three-dimensional entangled state}
\begin{figure}[bph]
\centering
\includegraphics[scale=0.14]{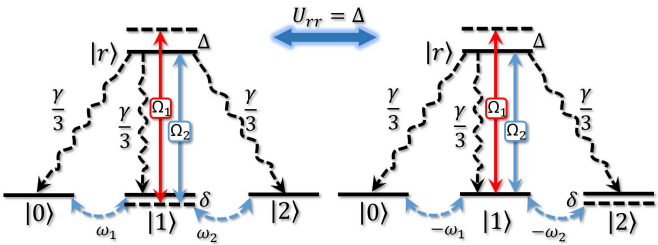}
\caption{The atomic level configuration of the scheme to prepare the three-dimensional entangled state with dissipation.}
\label{3Dmodel}
\end{figure}

\begin{figure}[bph]
\centering
\includegraphics[scale=0.52]{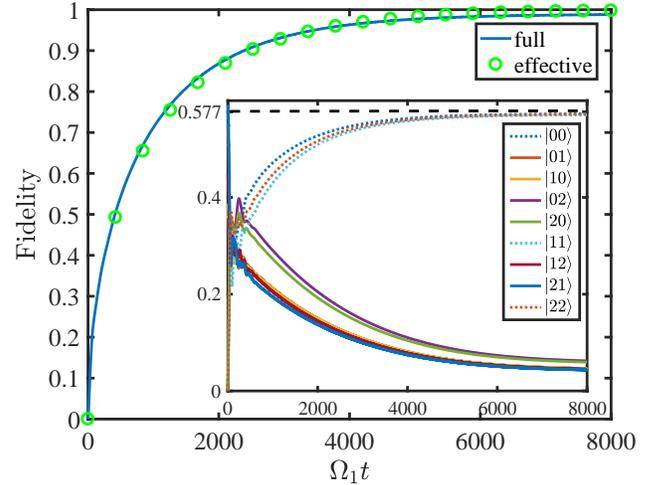}
\caption{The evolutions of fidelity of the three-dimensional state governed by the full (solid line) and effective master equation (empty circles). The inset shows the fidelities of the bared states.  The initial states are randomly chosen as $\rho_0=0.15|10\rangle\langle10|+0.35|21\rangle\langle21|+0.3|01\rangle\langle01|+0.2|12\rangle\langle12|$. The relative parameters are chosen as $\Omega_2=0.02\Omega_1,\omega_{1(2)}=0.01\Omega_1,\Delta=100\Omega_1$ and $\gamma=0.05\Omega_1$.}
\label{3DF}
\end{figure}
As is well known, the high-dimensional entanglement can not only  violate the local realism more strongly than the two-dimensional entanglement \cite{PhysRevLett.85.4418}, but also enhance the security of quantum key distribution \cite{PhysRevA.67.012311,PhysRevA.69.032313}. {Compared with the previous methods to generate the three-dimensional entanglement such as \cite{PhysRevA.92.022328} and \cite{pra012319}, we can acquire the three-dimensional entanglement with two identical atoms and fewer driving fields.} Once we adjust the dissipative scheme of two-dimensional entanglement slightly, the three-dimensional entanglement $|T_1\rangle=(|00\rangle+|11\rangle+|22\rangle)/\sqrt{3}$ can be prepared via URP and atomic spontaneous emission without specific initial state.

The scheme of three-dimensional entanglement includes two four-level Rydberg atom both consisting of three ground state $|0\rangle,|1\rangle,|2\rangle$ and one Rydberg state $|r\rangle$, which has been plotted in Fig.~\ref{3Dmodel}. For the first Rydberg atom, the ground state $|1\rangle$ is driven to the Rydberg state $|r\rangle$ by two independent laser fields with Rabi frequencies $\Omega_1$ and $\Omega_2$, detuning $-\Delta-\delta$ and $-\delta$, respectively. Meanwhile, it is coupled with the other ground states $|0\rangle$ and $|2\rangle$ with a resonant microwave field (Rabi frequency $\omega_1$) and a dispersive microwave field (Rabi frequency $\omega_2$, detuning $-\delta$), respectively. For the second Rydberg atom, the transition $|1\rangle\leftrightarrow|r\rangle$ is achieved by a dispersive laser field with Rabi frequencies $\Omega_1$, detuning $-\Delta$ and a resonant laser field with Rabi frequencies $\Omega_2$, respectively. The transitions between ground states $|0\rangle\leftrightarrow|1\rangle$ and $|1\rangle\leftrightarrow|2\rangle$ are resonantly and dispersively coupled by two microwave fields with Rabi frequencies $-\omega_1$ and $-\omega_2$, respectively. The detuning of the latter is $\delta$. The Hamiltonian of the total system can be written as
\begin{eqnarray}\label{3DH}
H_I&=&H_{R}+H_{MW},\\
H_{R}&=&\sum_{i=1}^2(\Omega_1e^{-i\Delta t}+\Omega_2)|r\rangle_i\langle1|+{\rm H.c.}+U|rr\rangle\langle rr|,\nonumber\\
H_{MW}&=&\sum_{i=1}^2(-1)^{i-1}\omega_1|1\rangle_i\langle 0|+(-1)^{i-1}\omega_2|1\rangle_i\langle 2|+{\rm H.c.}\nonumber\\
&&+\delta(|0\rangle_1\langle0|+|1\rangle_1\langle1|+|2\rangle_2\langle2|).\nonumber
\end{eqnarray}
The atomic spontaneous emission of the $i$-th atom can be described as $L_i^0=\sqrt{\gamma/3}|0\rangle_i\langle r|,~L_i^1=\sqrt{\gamma/3}|1\rangle_i\langle r|$ and $L_i^2=\sqrt{\gamma/3}|2\rangle_i\langle r|$. Then the form of the full master equation is similar to the Eq.~(\ref{2Dmaster}) with the new range of $k=0,1,2$.
Utilizing the URP, we can derive the effective Hamiltonian as
\begin{eqnarray}\label{3DeffH}
H_{\rm eff}&=&H^{R}_{\rm eff}+H^{MW}_{\rm eff},\\
H^{R}_{\rm eff}&=&\Omega_2(|10\rangle\langle r0|+|01\rangle\langle 0r|+|12\rangle\langle r2|+|21\rangle\langle 2r|)\nonumber\\
&&+{\rm H.c.},\nonumber\\
H^{MW}_{\rm eff}&=&\omega_1(|11\rangle-|00\rangle)(\langle01|-\langle10|)+\omega_1(|02\rangle\langle12|\nonumber\\
&&-|20\rangle\langle21|)+\omega_2(|11\rangle-|22\rangle)(\langle21|-\langle12|)\nonumber\\
&&+\omega_2(|10\rangle\langle20|-|01\rangle\langle02|)+{\rm H.c.}\nonumber\\
&&+\delta(|00\rangle\langle00|+|11\rangle\langle11|+|22\rangle\langle22|+|10\rangle\langle10|\nonumber\\
&&+|01\rangle\langle01|+2|12\rangle\langle12|+2|02\rangle\langle02|).\nonumber
\end{eqnarray}
The effective master equation with the corresponding Lindblad operators can be obtained as
\begin{eqnarray}\label{3DeffM}
\dot\rho&=&-i[H_{\rm eff},\rho]+\mathcal{L_{\rm eff}}\rho,\\
\mathcal{L_{\rm eff}}\rho&=&\sum_{k=1}^6L_{\rm eff}^k\rho L_{\rm eff}^{k\dag}-\frac{1}{2}(L_{\rm eff}^{k\dag}L_{\rm eff}^k\rho+\rho L_{\rm eff}^{k\dag}L_{\rm eff}^k),\nonumber
\end{eqnarray}
and
\begin{eqnarray}
L_{\rm eff}^1=\sqrt{\frac{\gamma}{3}}|00\rangle\langle0r|,\  \  L_{\rm eff}^2=\sqrt{\frac{\gamma}{3}}|01\rangle\langle0r|,\nonumber\\
L_{\rm eff}^3=\sqrt{\frac{\gamma}{3}}|02\rangle\langle0r|,\  \
L_{\rm eff}^4=\sqrt{\frac{\gamma}{3}}|00\rangle\langle r0|,\nonumber\\
L_{\rm eff}^5=\sqrt{\frac{\gamma}{3}}|10\rangle\langle r0|,\  \  L_{\rm eff}^6=\sqrt{\frac{\gamma}{3}}|20\rangle\langle r0|.\nonumber
\end{eqnarray}
According to the Eq.~(\ref{3DeffM}), we can notice that the ground states $\{ |00\rangle,|11\rangle,|22\rangle,|01\rangle,|10\rangle,|02\rangle,|20\rangle,|12\rangle,|21\rangle  \}$ are coupled with each other by the microwave fields, among which $\{ |01\rangle,|10\rangle,|12\rangle,|21\rangle  \}$ can be pumped into the excited states $\{|0r\rangle,|r0\rangle,|r2\rangle,|2r\rangle  \}$ by the laser fields. And these excited states will further decay to ground states via atomic spontaneous emission. Hence the system will repeat the processes
of pumping and decaying. However, in the absence of $\delta$, there are two steady states in the system, $|T_1\rangle$ and $|T_2\rangle=(3 |20\rangle+3|02\rangle+2|11\rangle-|00\rangle-|22\rangle)/2\sqrt{6}$. So we introduce the $\delta$ to filter the state $|T_2\rangle$ and turn the target state $|T_1\rangle$ into the unique steady state of the system, which means the system will be stabilized at $|T_1\rangle$ without a specific initial state.

In Fig.~\ref{3DF}, the validity of the reduced system has been verified due to the uniform behavior of evolutions of the fidelity  respectively governed by the full (solid line) and effective (empty circles) master equation. It is significant that the evolution of the fidelity of $|T_1\rangle$ reaches $98.8\%$ at $t=8000/\Omega_1$ with a random initial state. The inset shows the fidelities of the bared states, where the fidelities of states $|00\rangle,~|11\rangle$ and $|22\rangle$ (dotted lines) can all reach $0.572$ (the ideal values denoted by dashed line are $1/\sqrt{3}\simeq0.577$). Then, we also investigate the feasibility in experiment. The experimental parameters are set as $(\Omega_1,\gamma)=2\pi\times(1,0.03)$ MHz, $\Omega_2=0.02\Omega_1,~\omega_{1(2)}=0.01\Omega_1,$ and $\Delta=100\Omega_1$. The corresponding fidelity of the target state can be above $99.14\%$. All above results exhibit the reliability of the scheme for three-dimensional entanglement.

\section{autonomous quantum error correction}\label{V}

Quantum error correction has played an important role in the operation of quantum information processing, which is useful to protect the quantum computations from the quantum errors arising from uncontrolled interactions between the physical qubits and their
environment \cite{PhysRevA.52.R2493,PhysRevLett.77.793,PhysRevLett.77.3260}. Subsequently, the ingenious union of the quantum error correction and quantum dissipation was put forward by considerable ideas in theory and experiment \cite{ProcRSocLondA.454.355,PhysRevA.65.042301,PhysRevX.4.041039,Sci.347.853,PhysRevA.91.042322,PhysRevA.91.062324,PhysRevLett.116.150501,PhysRevA.96.012311,PhysRevB.95.134501,s41467}.
Most recently, Reiter \textit{et al.} presented an autonomous quantum error correction scheme with trapped ions \cite{s41467}, which inspires us to extend the URP to produce an autonomous quantum error correction with dissipation.

A logical qubit encoded in three physical qubits has a general form $|\psi\rangle=\alpha|0\rangle_L+\beta|1\rangle_L=\alpha|000\rangle_P+\beta|111\rangle_p$ with $|\alpha|^2+|\beta|^2=1$, where the subscripts $L$ and $P$ denote logical and physical, respectively. The states $|0\rangle_L$ and $|1\rangle_L$ are the basis states for the codespace. We can exploit the quantum error correction to protect the qubit $|\psi\rangle$ from being converted to the single-error state $|\psi_i\rangle=\sigma_x^{i}|\psi\rangle$, where $\sigma_x^{i}=|0\rangle_i\langle1|+|1\rangle_i\langle0|$ is the bit-flip error on the $i$-th physical qubit. The bit-flip noise can be described by $L_x^i=\sqrt{\Gamma}\sigma_x^i,~i=1,2,3$ with the bit-flip rate $\Gamma$. The corresponding master equation can reflect the noisy dynamics,
\begin{eqnarray}\label{noise}
\dot\rho&=&\mathcal{L}_{\rm noise}\rho\nonumber\\
&=&\sum_{i=1}^3L_x^i\rho L_x^{i\dag}-\frac{1}{2}(L_x^{i\dag}L_x^i\rho+\rho L_x^{i\dag}L_x^i).
\end{eqnarray}
In the following, we illustrate how to autonomously correct the bit-flip error via the URP in a Rydberg-atom-cavity system.
\begin{figure}
\centering
\includegraphics[scale=0.14]{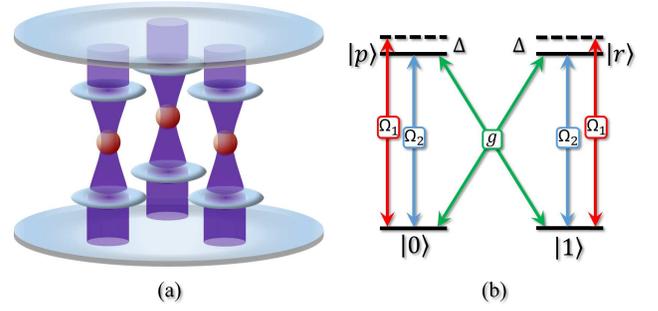}
\caption{(a) The setup for the autonomous quantum error correction. (b) The atomic level configuration of the scheme.}
\label{cmodel}
\end{figure}
In Fig.~\ref{cmodel}(a), we plot the setup for the autonomous quantum error correction, where three four-level Rydberg atoms are trapped in three independent optical cavities constructing an equilateral triangle to make the Rydberg interactions identical. The associated atomic level is shown in Fig.~\ref{cmodel}(b), which consists of two ground states $|0\rangle$ and $|1\rangle$, and two Rydberg states $|p\rangle$ and $|r\rangle$. The ground state $|0(1)\rangle$ is used as encoded quantum bit and is coupled with the Rydberg state $|p(r)\rangle$ by a dispersive laser field (frequency $\Omega_1$, detuning $\Delta$) and a resonant laser field ( Rabi frequency $\Omega_2$). Simultaneously, the transition between $|0(1)\rangle$ to $|r(p)\rangle$ is driven by the quantized cavity field resonantly with
coupling strength $g$. In the interaction picture, the Hamiltonian reads
\begin{eqnarray}
H_I=H_L+H_Q,
\end{eqnarray}
and
\begin{eqnarray}
H_L&=&\sum_{i=1}^3(\Omega_1e^{-i\Delta t}+\Omega_2)(|p\rangle_i\langle0|+|r\rangle_i\langle1|)+{\rm H.c.}\nonumber\\
&&+\sum_{j>i}U_{rr}^{ij}|rr\rangle_{ij}\langle rr|+U_{pp}^{ij}|pp\rangle_{ij}\langle pp|\nonumber\\
&&+U_{rp}^{ij}|rp\rangle_{ij}\langle rp|+U_{pr}^{ij}|pr\rangle_{ij}\langle pr|,\nonumber\\
H_Q&=&\sum_{i=1}^3g(|p\rangle_i\langle1|+|r\rangle_i\langle0|)a_i+{\rm H.c.},\nonumber
\end{eqnarray}
where $U_{\alpha\beta}^{ij}$ stands for the Rydberg interaction between the $i$-th atom in $|\alpha\rangle$ and the $j$-th atom in $|\beta\rangle$, and $a_i$ denotes the annihilation operator of the $i$-th cavity. Once we select the suitable principle quantum numbers of Rydberg atoms \cite{PhysRevA.77.032723,PhysRevA.91.043802}, the Rydberg interactions can be considered as $U_{rr}=U_{rr}^{ij}=U_{pp}^{ij}\gg U_{rp}^{ij}=U_{pr}^{ij}$. Then we can simplify the $H_L$ as
\begin{eqnarray}
H_{\rm full}&=&\sum_{i=1}^3(\Omega_1e^{-i\Delta t}+\Omega_2)(|p\rangle_i\langle0|+|r\rangle_i\langle1|)+{\rm H.c.}\nonumber\\
&&+\sum_{j>i}U_{rr}(|rr\rangle_{ij}\langle rr|+|pp\rangle_{ij}\langle pp|).
\end{eqnarray}
The dissipative process of the $i$-th cavity can be described as $L_c^{i}=\sqrt{\kappa}a_i,$ where $\kappa$ is the decay rate of cavity. When we consider $\kappa\gg g$ to adiabatically eliminate $a_i$ \cite{PhysRevA.85.032111}, the interaction between the $i$-th quantized cavity field and atom can be equivalent to a Lindblad operator $L_e^{i}=\sqrt{\kappa_e}(|0\rangle_i\langle r|+|1\rangle_i\langle p|)$, where $\kappa_e=4g^2/\kappa$. The dynamics of three-atom system can be described by a master equation
\begin{eqnarray}\label{cfullM}
\dot\rho&=&-i[H_{\rm full},\rho]+\mathcal{L}_e\rho,\\
\mathcal{L}_e\rho&=&\sum_{i=1}^3L_e^i\rho L_e^{i\dag}-\frac{1}{2}(L_e^{i\dag}L_e^i\rho+\rho L_e^{i\dag}L_e^i).\nonumber
\end{eqnarray}
\begin{figure}
\centering
\includegraphics[scale=0.52]{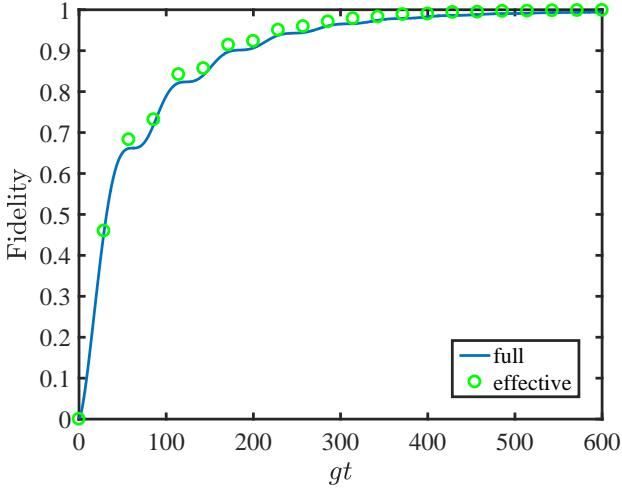}
\caption{The fidelity of state $(|000\rangle+i|111\rangle)/\sqrt{2}$ as a function of $gt$ with Eq.~(\ref{cfullM}) (solid line) and Eq.~(\ref{ceffM}) (empty circles). The initial state is $(|100\rangle+i|011\rangle)/\sqrt{2}$ and the other parameters are $\Omega_1=3g,~\Omega_2=0.05g,~U_{rr}=\Delta=800g$ and $\kappa_e=0.02g$.}
\label{cF}
\end{figure}
\begin{figure}[bph]
\centering
\includegraphics[scale=0.52]{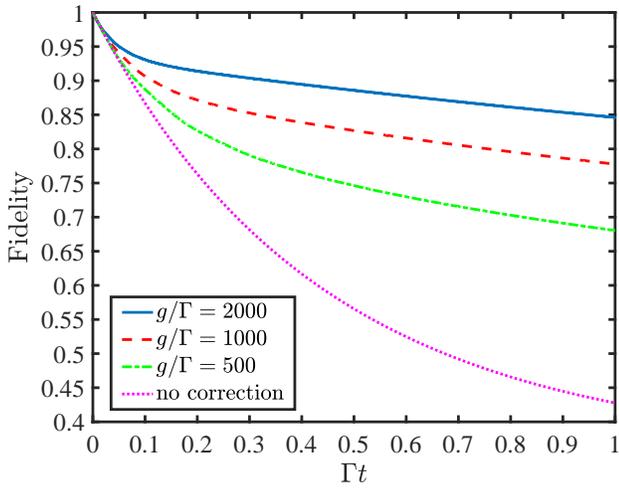}
\caption{The evolution of the fidelity of state $(|000\rangle+i|111\rangle)/\sqrt{2}$ governed by the Eq.~(\ref{totM}) with different coupling strength $g$. The initial state is $|\psi(0)\rangle=(|000\rangle+i|111\rangle)/\sqrt{2}$ and the relevant parameters are: $\Omega_1=3g,~\Omega_2=0.05g,~U_{rr}=\Delta=800g$ and $\kappa_e=0.02g$.}
\label{crate}
\end{figure}
In the condition of URP, $U_{rr}=\Delta\gg\Omega_1\gg\Omega_2$, the evolution of reduced system will be governed by
\begin{eqnarray}\label{ceffM}
\dot\rho&=&-i[H_{\rm eff},\rho]+\mathcal{L}_e\rho,\\
H_{\rm eff}&=&\Omega_2\big(|100\rangle\langle r00|+|010\rangle\langle 0r0|+|001\rangle\langle00r|\nonumber\\
&&+|110\rangle\langle 00p|+|101\rangle\langle 0p0|+|011\rangle\langle p00|\big)\nonumber\\
&&+{\rm H.c.}.\nonumber
\end{eqnarray}

One can find that our scheme can correct the single-error states by pump them to the single excited states, which further decay to the desired stable states $|000\rangle$ or $|111\rangle$ via leaky cavities. In order to visually reveal the correcting process, we perform how the three-atom system evolves when a single-error state $(|100\rangle+i|011\rangle)/\sqrt{2}$ occurs, which is characterized by the fidelity of state $(|000\rangle+i|111\rangle)/\sqrt{2}$ as a function of $gt$ with Eq.~(\ref{cfullM}) (solid line) and Eq.~(\ref{ceffM}) (empty circles) in Fig.~\ref{cF}.
It is clear enough that the two curves are in good agreement with each other and a high fidelity can be up to $99.6\%$ at $gt=1000$. The appearance affirms the feasibility of our quantum error correction scheme and the correctness of the reduced system.
In experiment, the transition $|0\rangle\leftrightarrow|r\rangle$ requires two indirect transitions to realize \cite{PhysRevA.95.022317}. Firstly, the ground state $|0\rangle$ dispersively coupled with a intermediate state $|e\rangle$ by an optical cavity with strength $g_b$, detuning $-\Delta_b$. Secondly, the intermediate state $|e\rangle$ will be pumped to the Rydberg state $|r\rangle$ via a classical field with Rabi frequency $\Omega_b$, detuning $\Delta_b$. In the regime of the large detuning, $|\Delta_b|\gg \{g_b,\Omega_b\}$, the intermediate state $|e\rangle$ can be eliminated adiabatically and an equivalent direct transition $|0\rangle\leftrightarrow|r\rangle$ can be accomplished with an effective strength $g_{\rm eff}=g_b\Omega_b/\Delta_b$, which is analogous to the strength $g$ in our scheme. Consequently, we can experimentally regulate the value of $\Delta_b$ to obtain desired values of $g$ and $\kappa_e$. Then we substitute a group of experimental parameters $(\Omega_1,\Omega_2,\Delta,\gamma,\kappa_e)=2\pi\times(3,0.05,800,0.001,0.02)$ MHz and the fidelity of state $(|000\rangle+i|111\rangle)/\sqrt{2}$ can be above $97.34\%$.

In Fig.~\ref{crate}, we analyze the capability of autonomously correcting the single-error state as the bit-flip noise continuously emerges. The total process of autonomous quantum error correction reads
\begin{eqnarray}\label{totM}
\dot\rho&=&-i[H_{\rm full},\rho]+\mathcal{L}_e\rho+\mathcal{L}_{\rm noise}\rho.
\end{eqnarray}
When the error correction is absent (dotted line), the fidelity will steeply descend to $42.77\%$ at $\Gamma t=1$. Nevertheless, as long as the error correction arises, the bit-flip noise can be autonomously and continuously corrected, and the decline of fidelity will be repressed remarkably with the enhancement of $g$. At $\Gamma t=1$, the fidelity can reach $68.05\%,~77.77\%$ and $84.62\%$ with $g=500\Gamma$ (dotted-dashed line), $g=1000\Gamma$ (dashed line) and $g=2000\Gamma$ (solid line), respectively.

\section{Summary}\label{VI}
In summary, we have successfully realized a new mechanism, unconventional Rydberg pumping (URP), via the organic combination of Rydberg interaction and two classical fields. We can take advantage of the URP to freeze the evolution of the states with two atoms at the same ground state, and excite the states with two atoms at different ground states. Then we apply the URP to actualize the three-qubit controlled phase gate, the two- and three-dimensional steady-state entanglement and the autonomous quantum error correction. The corresponding results can adequately evidence the feasibility of all above applications by considering the state-of-the-art technology. We believe our scheme supplies a new prospect on quantum information processing with neutral atoms.

\section*{ACKNOWLEDGMENTS}
This work is supported by National Natural Science Foundation of China (NSFC) under Grants No. 11774047.

\bibliography{URP}

\end{document}